# Study of the Sorption Properties of $Ge_{20}Se_{80}$ Thin Films for $NO_2$ Gas Sensing

Ping Chen[a], Maria Mitkova[a,*], Dmitri A. Tenne[b], Kasandra Wolf[a], Velichka Georgieva[c] and Lazar Vergov[c]

[a]Dept. of Electrical and Computer Engineering, Boise State University, Boise, ID 83725-2075

[b]Dept. of Physics, Boise state University, Boise, ID 83725-1570, USA

[c]Laboratory of Acoustoelectronics, Georgi Nadjakov Institute of Solid State Physics (ISSP), Bulgarian Academy of Sciences (BAS), Bulgaria

Abstract

In this study we investigated the sorption ability of $Ge_{20}Se_{80}$ thin films applied as active layers of quartz crystal microbalance (QCM) for $NO_2$ gas sensing. To identify the chalcogenide system appropriate for gas sensing, we provided data for the packing fraction of a number of chalcogenide systems and discussed their suitability. We performed Raman spectroscopy, X-ray photoelectron spectroscopy (XPS) and atom force microscopy (AFM) measurements on the thin films both before and after gas absorption, which showed that the introduced gas molecules interact electrostatically with the chalcogen atoms of the host material and initiate some degree of structural changes in it. The weight change due to $NO_2$ gas absorption was measured by frequency change of the resonator. The absorbed mass increased monotonically with the thickness of chalcogenide films and the $NO_2$ gas concentration. At the conditions of our experiment, up to 11.4 ng of the gas was absorbed into 200nm thick $Ge_{20}Se_{80}$ film at 5000 ppm $NO_2$ concentration. The process of gas molecules absorption is found irreversible at the purging conditions.

[*] Corresponding author at: Dept. of Electrical and Computer Engineering, Boise State University, Boise, ID 83725-2075, USA. Tel.: +1 208 4261319; fax: +1 208 4262470.
E-mail address: mariamitkova@boisestate.edu (M. Mitkova).

## 1. Introduction

Chalcogenide glasses attracted widespread attention as material system for information storage [1, 2], optical recording [3, 4] and opto-mechanical effects [5]. One recently emerging function of these materials is their inclusion as an active film in $NO_2$ gas sensing [6, 7], which has proven to offer reasonable sorption and desorption ability in the initial experiments. We specifically mention $NO_2$ and focused our work on this gas because it is a toxic gas, whose concentration in the air dramatically increased lately, released by combustion, automobiles and some plants. Its concentration in the air affects human health, flora and fauna, the quality of water and soil, i.e. the ecosystem as a whole. This imposes a strict monitoring of the $NO_2$ quantity in the environment, for which different sensing systems have been developed. Some of them are based on principle of measuring the electrical response of the sensors [8]. These devices are very compact and cheap but they have low sensitivity [9], a long response time [10] or operate at high temperatures [11]. There is one very interesting idea to use surface plasmon resonance in chalcogenide glasses for sensing [12], which has been theoretically studied. The model created through that study promises good sensitivity and is feasible in the entire IR region, which makes it competitive in the field of gas sensing, though it is not realized in an experimental environment.

The method of gas sensing we chose to study the sorption ability of $Ge_{20}Se_{80}$ films is based on the thin-film - quartz crystal microbalance (QCM) system [13]. It surpasses the performance of all above mentioned methods, is relatively cheap and presents the unique property of QCM to detect mass in nano-gram scale. Other attributes of such a system are an absolute measurement precision value ± 0.1 Hz, applicability in a wide concentration range from ppm to ppb, independence of the signal from strong electric, magnetic and radiation fields, a suitable initial signal for digital processing of the information, and long-term stability work of the elements. The key for the successful application of this method is the choice of the thin film, which comes in contact with and absorbs the gaseous molecules. In order to be competitive towards the other known materials, the absorbing films must guarantee a minimum deterioration of the resonator's parameters due to the processes of layer deposition and allow a maximum mass loading of $NO_2$. Currently the most matured in this respect is the technology of inclusion of thin oxide films – $MoO_3$ and $WO_3$ [14]. It offers good sensitivity and response but requires additional procedures for formation of well organized crystalline structure of the films and high operating temperature. The latter is indeed good for some niche applications, like the automotive industry. In the majority of cases however, operation at room temperature is desirable where the use of those oxide films is impossible.

Since all studies of chalcogenide glasses as an absorption medium for gas sensing are pretty much sporadic and not based on a preliminary global assessment of their performance, in the current work we present some theoretical considerations of what would be important in the choice of the gas absorbing film material based on chalcogenide glasses and then report our data related to absorption of $NO_2$ in a QCM with $Ge_{20}Se_{80}$ active film.

## 2. Theoretical considerations

There are two major requirements towards the material of the active film for a QCM based gas sensor. It has to possess low density and a very open structure in order to absorb the gas molecules easily and consists of light atoms, since the sorption sensitivity increases with the thickness of the films. That means that the films must be light enough in order to not destroy the function of the resonator. In other words, it is desirable to create active films as thick as possible, which are expected to absorb more gaseous molecules and still assure good resonance parameters of the QCM. The chalcogenide glasses have already been a subject of such studies [15] which revealed their possible applicability in the field of gas sensing and the physical nature of the sorption process.

One initial suggestion for the gas absorption abilities of chalcogenide glasses arises when we deduce the packing fractions from their densities and molar volumes. The packing fraction will provide a dimensionless parameter that displays some universal trends about how dense or porous the material is. The packing fraction is given by

$$pf = \Sigma(4/3\pi\, r(i)^3\, n(i)\, N_A / V_m), \qquad (1)$$

where $r(i)$ is the ionic radius of the $i^{th}$ atom, $n(i)$ is the mole fraction of that corresponding atom presented in the structural formula, $N_A$ is the Avogadro constant and $V_m$ is the molar volume. Packing fraction for thermodynamically stable binary chalcogenide glasses and some Ag-containing chalcohalide glasses are shown in Fig. 1. The correct choice of ionic radii is important here because it is the main uncertainty in the packing fraction calculation. For Ag and I atoms, reported ionic radii corrected for coordination number from Shannon [16] were used. While for covalent bonded P, S, Ge, As, Se and Te atoms, consistent ionic radii data are not available and empirical covalent radii [17] were used to ensure that the relative trends are meaningful. The molar volume data are taken from Ref. [18-21].

We would expect materials with lower packing fractions to be good hosts for sorption of $NO_2$ molecules. As shown in Fig. 1, the sulfur or phosphorus rich glasses tend to have low packing fractions because sulfur or phosphorous monomers cause high vapor pressure at quench temperature and the resulting voids in the glasses would decrease the packing fractions. However, those are not good gas sensing candidates due to the glass inhomogeneity and instability. For AgI or Te containing glasses, the packing fractions are higher than glasses comprised of only Ge, As or Se atoms. This is due to a severe size mismatch of atoms in the former type of glasses where smaller atoms can fill up interstitial sites between larger atoms and increase the packing fraction, which reduces the absorption ability of these glasses.

Another quantity to characterize the packing in different glass systems is the free volume, which is the volume increase of the glass compared to its crystalline counterparts. The larger free-volume would correspond to lower packing fraction. For example, free volume in the Ge-Se glasses is around 10-15 % of the overall volume with the free volume maximum occurring at $Ge_{33}Se_{67}$ composition [19], which coincides with the minimum in packing fraction of Ge-Se glasses in Fig. 1. At the minimum packing glass composition, a noticeable amount of volume will be empty and this empty space will aid in the migration of gaseous molecules through the material. By varying the size of the chalcogen atom from S to Se and Te, the bond angles and structure will concurrently change and cause a changing void size[22, 23]. One of the most accepted models on voids is the Cluster Bypass Model. In the original Cluster Bypass Model, clusters of high-density occur during glass transition [22, 24]. The clusters grow to a particular size until their self-limiting nature prevents further expansion. Material not included in the clusters creates a low-density, connective tissue where the gaseous molecules could migrate. Consequently, the free volume consists of channels and some closed voids. Such morphology would greatly depend on the choice of higher coordinated additive atoms in the chalcogenide glasses.

With these considerations in mind, Ge-Se glass system seems to be the best candidate for gas sensing applications. It is a good glass former in wide composition range, porous with low packing fraction, nontoxic unlike As containing glasses. As shown in Fig. 1, there are two packing fraction minima in Ge-Se system. One is at stoichiometric $GeSe_2$ composition and the other resides at chalcogen-rich side including pure selenium glass. The former has been studied by us which indeed showed good reversible sorption ability [15]. For the latter, the very chalcogen-rich glasses are not suitable since those are subject to partial crystallization and phase separation. Because of this a study of the $Ge_{20}Se_{80}$ glasses presents a special interest and is the subject of this work. Based on the rigidity transition theory, the optimum value of mean coordination $r = 2.4$ has been introduced at which the glass condition is most stable. The quantity $r$ is formally defined as $(\Sigma\, n_i r_i)/\Sigma n_i$, where $n_i$ represents the number of atoms having a coordination $r_i$ and $\Sigma n_i$ gives the total number of atoms in a network [25]. Its importance is well appreciated in covalently bonded systems [26]. In respect to these findings, the $Ge_{20}Se_{80}$ composition with $r = 2.4$ has a special position within the Ge-Se glass forming region due to its high glass forming tendency and self-origination [27] while still maintaining relatively low packing fraction and a large free volume. These considerations were a strong motivation to study the $Ge_{20}Se_{80}$ glass films as a possible self-organized and non-aging absorption medium. We characterized the film itself and the changes occurring in it as a result of the absorption of $NO_2$ molecules and provided data of its performance in a quartz resonator system.

## 3. Experimental

16 MHz QCM's with 4 mm diameter gold electrodes were created on AT-cut quartz wafers. $Ge_{20}Se_{80}$ films were evaporated onto both sides of QCM's under high vacuum with four different thicknesses (50nm, 100nm, 150nm and 200nm). These resonators were then strictly stored in vacuum to avoid interaction with humidity or other gases in ambient air until $NO_2$ gas absorption experiments were performed. The as-deposited films were studied by energy-dispersive X-ray spectroscopy (EDS), Raman spectroscopy, grazing angle X-ray diffraction (GAXRD), X-ray photoelectron spectroscopy (XPS) and atomic force microscopy (AFM), in order to define their chemical composition, structure and surface morphology. All these characterization methods were performed in vacuum or controlled environment not containing $NO_2$.

EDS was done on a LEO 1430VP Scanning Electron Microscope with EDS accessory to determine the elemental composition of the thin films. Since a semi-Knudsen cell evaporator has been used for the films deposition of the samples, their composition was very close to the source composition. The EDS data are presented in Table. 1.

Raman spectra were measured using a Horiba Jobin Yvon T64000 triple monochromator equipped with a liquid-nitrogen-cooled multichannel coupled-charge-device detector. The samples were excited with the 441.6 nm line of a He-Cd laser and the power on sample was 60 mW focused into a circular spot of ~ 0.2 mm in diameter. The sample chamber was pumped down to $1x10^{-5}$ Torr to avoid oxidation and the samples were cooled to 100K during Raman measurements to reduce the chance for occurrence of photo-induced effects due to laser irradiation. For these thin film samples, we did not see any photo darkening effects under microscope after the Raman laser irradiation and the line shapes remained the same over time. Thus, we are confident that the conditions used for Raman experiments (the above mentioned laser power density and cooling the samples down to 100K in vacuum) are appropriate for obtaining reproducible results without causing additional light induced effects. After collection of the Raman data, a deconvolution of all normalized Raman spectra was performed (the normalization was carried out after a preliminary subtraction of a baseline).

Grazing XRD measurements were made on a Bruker AXS D8 Discover X-Ray Diffractometer with Cu $K\alpha_1$ radiation ($\lambda$=1.5406 Å) using a grazing (1° or 3°) incidence geometry over 4° to 100° in 2θ (step size is 0.1°and 6 sec/step) on a NaI(Tl) scintillation detector which helped for exact evaluation of the bond lengths in the structure of the films.

The X-ray photoelectron spectroscopy (XPS) was measured on a Physical Electronics Versaprobe. Samples were irradiated with a monochromated Al $K_\alpha$ x-ray beam approximately 100 μm in diameter at about 100 watts scanned over a 1.4 mm x 0.1 mm area. The film samples were mounted on a sample stage by securing the edges with washers and screws. The spectrometer pass energy was set at 117.5 eV for the survey scan and 46.95 eV for the high resolution spectra, and the binding energy scale was calibrated using the Cu $2p_{3/2}$ and Au $4f_{7/2}$ peaks from freshly sputter cleaned 99.9% pure Cu and Au foils (Alfa Aesar). The spectrometer acceptance window was oriented for a take-off angle of 45° from the sample normal. These conditions produce full width at half max of better than 0.92 eV for Ag $3d_{5/2}$. To minimize sample charging, low energy electrons and Ar ions bleeding over the sample was applied.

The surface roughness and morphology was studied by atomic force microscopy (AFM) with Veeco Dimension 3100 Scanning Probe Microscope with a Nanoscope V controller.

A schematic diagram of the gas absorption experimental set up is presented in Fig. 2. The first step was to measure the equivalent dynamic parameters of the resonators to be sure that they kept their parameters after the process of films depositions. All devices were in good condition and ready for the absorption experiments. The set up contained the following basic modules: a gas module (GM) – bottles with carrier gas, purge gas and test gas; a gas mix and control module (GMCM) – which included two mass flow controllers (FC-260 and FC-280) and a mixing camera; a test chamber (TC) with a Pt-thermosensor (PS) and a mass sensitive sensor (MS); a thermostat module (TM); a generator and frequency counter (GFC) and a computer system (CS).

The QCM was installed on a special holder inside the test chamber. The temperature of the sample was measured by a Pt-thermosensor positioned near the QCM. First, the chamber was air-scavenged, and then test gas with a certain concentration was released as a permanent flow. The velocities of both the carrier and test gases were measured and controlled by mass flow controllers, their ratio being defined by the desired concentration.

A frequency counter (Hameg 8123) connected to the QCM as well as to the computer for data recording registered the QCM frequency. In this way, the frequency change as a function of time was identified. The initial frequency value was measured under the saturated carrier gas flow conditions. The gas to be tested came from certified bottles diluted with synthetic air. The test gas was continuously added to the carrier gas to obtain the desired test gas concentration. After adding the mixtures of gases into the system, the frequency of the QCM started to decrease. After a certain period of time it reached a constant value, when a dynamic equilibrium at certain gas concentration and temperature was established. A temperature of 28.6$^{o}$C was maintained in the test chamber. Experiments with $NO_2$ concentrations in the gas flow from 100 to 5000 ppm were carried out. At the end of the experiments, the pollutant flow was terminated and the frequency was measured to see if there was any increase as a result of desorption of mass from the QCM. At that stage, the measurement was finished.

## 4. Results

### *4.1 Raman scattering analysis*

The purpose of the Raman investigation was to obtain information on absorption-induced changes in the film structure. Inelastic light scattering is known to be sensitive to material structure (the type of the structural units, its connection and amount) thus giving rise to observations of some relative changes in the intensity of the vibration modes. Spectra of virgin films and $NO_2$ gas absorbed films were taken. Deconvolution of the measured Raman spectra was performed in order to distinguish the vibration modes having contribution in the integrated light scattering from each sample. The films had the expected Raman features appearing at the positions corresponding to those for bulk material with an identical composition, which is an indication that the films which were produced are relaxed and have structure identical to this of the bulk materials with the same composition. The deconvoluted Raman spectra of 200nm-thick deposited $Ge_{20}Se_{80}$ glass films before and after gas absorption are presented in Fig. 3. A slight increase in the area intensity ratio of the edge-sharing units *v.s.* corner-sharing units from 27.9% to 29.9% suggests some interaction between the hosting film and the $NO_2$ molecules.

*4.2 X-Ray diffraction studies*

The grazing XRD results are shown in Fig. 4. The plot shows a prepeak lying at 0.99-1.01 $Å^{-1}$ smaller than $Q_p$, the position of the principal peak of the diffraction pattern, which is determined by the nearest-neighbor distance $r_1$ in real space. This prepeak is the so-called 'first sharp diffraction peak' (FSDP) and it corresponds to real-space structural correlations on length scales appreciably larger than $r_1$, which is in the medium range order (MRO) range. The effective periodicity, $R$, can be related to the position of FSDP, $Q_1$ [28]

$$R \approx 2\pi / Q_1 \qquad (2)$$

The correlation length, $D$, over which such quasi-periodic real-space density fluctuations are maintained can be obtained from the full width at half maximum (FWHM), $\Delta Q_1$, of the FSDP using the expression [28]

$$D \approx 2\pi / \Delta Q_1 \qquad (3)$$

It has been proposed by Elliott that the FSDP can be represented by a structural model in which ordering of interstitial voids occurs in the structure [28]. Also, Blétry [29] has given a simple formula to relate FSDP to the cation-cation nearest-neighbor distance $d$ for $AX_2$ type materials, namely,

$$d \approx 3\pi / 2Q_1 \qquad (4)$$

Based on these equations, we obtained the following data for the studied films:

$Ge_{20}Se_{80}$: $Q_1 = 1.01$ Å$^{-1}$, $\Delta Q_1 = 0.30$ Å$^{-1}$, Effective periodicity, $R \approx 2\pi / Q_1 = 6.22$ Å;

Correlation lengths, $D \approx 2\pi / \Delta Q_1 = 20.9$ Å; cation-cation distance $d = 4.67$ Å

This result shows that the cation-cation distance is only 27% greater than the length of $NO_2$ molecule (3.4 Å). This suggests occurrence of a very intimate connection between the $NO_2$ molecules and the 4-member or 6-member rings formed in network building units which will be discussed in the next section.

## *4.3 X-ray photoelectron spectroscopy*

Fig. 5 shows Ge 3d core level XPS spectra of as-deposited $Ge_{20}Se_{80}$ film and the same batch film after $NO_2$ gas absorption. In Ge 3d core level XPS spectra, one can see an increase in the amount of Ge – O bonds after gas absorption, indicative of some oxidation by $NO_2$ gas. Besides, for the N 1s peak there is a clear increase of the band around 400 eV in $Ge_{20}Se_{80}$ composition after gas absorption as shown in Fig. 6. We assign the peak to atomic nitrogen [30]. This is a direct evidence of $NO_2$ sorption into the glass backbone. After 10 seconds of $Ar^+$ ion sputtering, a shoulder peak around 405 eV corresponding to physisorbed $NO_2$ occurs [30], which could be due to physisorption of released $NO_2$ molecules back to the thin film surface.

## *4.4 Atomic force microscopy*

Measurements were performed to get information about the surface morphology of the films. It can give an idea about the absorption ability of the films since this is related to the structure of the material and the free surface, which is bigger at rougher structures. There were AFM data collected from different points of all samples showing that the films are relatively smooth. The AFM scans for 200 nm thick film are shown in Fig. 7. The surface roughness $R_q$ increased from 1.0 nm for virgin samples to 2.1 nm after gas absorption while still maintaining similar surface morphology.

### *4.5 Sorption properties of $Ge_{20}Se_{80}$ thin films*

Fig. 8 shows the time-frequency characteristic of a 200nm thick $Ge_{20}Se_{80}$ thin film sample at a constant $NO_2$ concentration of 500 ppm. The measurement starts when the frequency of the synthetic air flow becomes constant. In the $NO_2$ flow the system is constantly powered in time till saturation at 2500 sec is reached. When purging synthetic air is forced through the system for 600 sec, the absorbed gas stays in the films, i.e. the process is *irreversible* at the purging conditions. The maximum change in the frequency is 15 Hz. This is indeed the difference between the initial frequency and the one at which saturation occurs, and corresponds to a weight increase of 3.4 ng.

The response of the chalcogenide films towards the concentration of the $NO_2$ gas was studied at gas phase concentration increased in steps (100, 500, 1000, 2500 and 5000 ppm), while additional waiting time was allowed at each step for the resonator to reach a saturated weight and frequency reading.  The $NO_2$ concentrations were labeled next to each saturation line. The total frequency-time characteristic is stepwise as shown in Fig. 9.

The maximum frequency change $\Delta f$ in the system at different $NO_2$ concentration conditions are calculated. According to the Sauerbrey equation [31] for AT-cut quartz resonators, the calculation of the absorbed mass $\Delta m$ for each experiment in nanograms from the measured frequency change $\Delta f$ is:

$$\Delta f = -2.26\ 10^{6} f_0^{2}\ \Delta m / s \quad (5)$$

where $\Delta f$ and $f_0$ are in Hz and MHz, $\Delta m$ is in gram and $s$ in $cm^2$ is the area of the QCM electrode.

The maximum frequency shift $\Delta f$ and absorbed mass $\Delta m$ at different $NO_2$ concentration steps for films with different thicknesses are combined together in Fig. 10. As illustrated, the frequency shift and sorbed mass increase with increased film thickness. For the 200nm thick film, the maximum weight increase is 11.4 ng at 5000 ppm $NO_2$ concentration which corresponds to $2.49\times10^{-10}$ mole of $NO_2$ molecules. Since the molar volume of $Ge_{20}Se_{80}$ bulk glass is 17.77 $cm^3$/mole [18], we can calculate that there is a total amount of $1.41\times10^{-7}$ mole of Ge and Se atoms in the 200nm thick film. That results in one absorbed $NO_2$ molecule in approximately every 568 atoms of the Ge-Se backbone.

## 5. Discussions

There are several simultaneously acting factors which contributed to the performance of the chalcogenide films as shown above.

In the first place, it is important to understand how the gaseous molecules which remain in the film after purging are prevented to escape by the chalcogenide film. There are not convincing evidences for occurrence of chemical reaction between the chalcogenide backbone and the $NO_2$ molecules, hence the later are partially held in the chalcogenide films by physisorption. These molecules could be released from the film during the outgasing experiments in case they can easily move through the structure of the chalcogenide glass. Note that the films also react relatively slow on absorption, the reason for which could be the reduced free volume in this particular glass compared to the $Ge_{33}Se_{67}$ counterpart that affects the general performance of the films. However, the fact that the films do not desorb the gaseous molecules in the particular time in which the system was purged, shows that there is a stronger bonding which makes the process irreversible. We suggest that the reason for the strong coupling of the $NO_2$ molecules with the glass structure is the specific electron structure of Se which is two-fold coordinated with two of its *p* electrons participating in a sigma bonding and the other two *p* electrons forming a lone pair. The presence of the lone pair electrons gives rise to a high negative effective correlation energy [32], which results in the transformation of two neutral chalcogen atoms $C_2^0$ to a pair of charged atoms with higher and lower coordination, ($C_1^-$ and $C_3^+$) respectively. If the $C_1^-$ atom comes in contact with the $NO_2$ molecule which has a positive charge positioned at the nitrogen atom – Fig. 11 [33], a strong electrostatic interaction occurs. This all affects the overall electronegativity of the system and it is because of this reason that we registered a shift of the binding energy of Ge in the XPS studies that is related to an increase of the number of Ge atoms in oxidized condition and attachment of the $NO_2$ molecules to the Se atoms. This evidently is also related to some quite limited reorganization of the structure in which, as revealed by the Raman studies, the number of the edge-shared building blocks slightly increases. This structural reorganization also contributes to the increase of the roughness of the surface as shown in the AFM studies.

The XRD study showed that the bond lengths in the structure of the hosting $Ge_{20}Se_{80}$ glass contribute to formation of a structure with large enough openings in which the $NO_2$ molecules can be accepted. In a macro aspect, the Ge-Se corner-sharing and edge-sharing tetrahedra form ring type of structures with 4 or 6 members as predicted by the *ab initio* calculations of D. Drabold et al. [34, 35] for very close in compositions chalcogenides. This structure is open enough in order for the $NO_2$ molecules to diffuse in it. Because of the electrostatic effects and slight structural reorganization, they cannot easily escape from the hosting material which makes the gas sensor based on this type of glasses non-reversible. A structural model corresponding to the initial step of the $NO_2$ diffusion into the $Ge_{30}Se_{70}$ glass is presented in Fig. 12. So, the sorption process of $NO_2$ in the $Ge_{20}Se_{80}$ is physical in nature but the forces occurrying at it are strong enough to prevent its reversibility. Compared to the $Ge_{33}Se_{67}$ [15] the studied films absorb lower amount of $NO_2$ which is in harmony with the reduced free volume predicted by the density and structural factor data.

## 6. Conclusions

The studies show that the absorption of $NO_2$ in $Ge_{20}Se_{80}$ has irreversible character. The main reason for this is the strong electrostatic attachment of the $NO_2$ molecule to the chalcogenide atoms and the occurrence of a structural reorganization within the chalcogenide glass. Evidence of the influence of the gaseous molecules on the structure of the films has been collected through Raman spectroscopy. This revealed an increase of the number of edge-sharing tetrahedra in the structure. Similarly, the AFM studies demonstrated increase of the surface roughness after absorption of the gaseous molecules. Besides, the XPS studies showed a slight increase of the Ge-O bonds after absorption as well as an increased presence of nitrogen, which most likely interacted with chalcogen atoms after absorption. All these specifics of the process with the studied glass films make the $NO_2$ absorption process non-reversible.

**Acknowledgements**

This work has been conducted as a collaboration between Boise State University and Bulgarian Academy of Sciences - Institute for Solid State Physics. Authors are thankful to the National Science Foundation (NSF) via IMI (International Materials Institute for New Functionality in Glass), Grant DMR 0844014. We would like to acknowledge Brian Jaques and Gordon Alanko for the help with XRD and XPS measurements, respectively. XRD and XPS instruments have been supported by NSF-Major Research Instrumentation awards No. 0619795 and 0722699.

## References

[1] M.N. Kozicki, M. Mitkova, in: R. Waser (Ed.), Nanotechnology: Volume 3: Information Technology I, Wiley-VCH Verlag GmbH & Co. KGaA, Weinheim, 2008.

[2] M. Wuttig, Nat Mater 4 (2005) 265.

[3] M. Kincl, L. Tichy, Mater. Chem. Phys. 103 (2007) 78.

[4] V. Lyubin, M. Klebanov, M. Mitkova, Applied Surface Science 154-155 (2000) 135.

[5] P. Krecmer, A.M. Moulin, R.J. Stephenson, T. Rayment, M.E. Welland, S.R. Elliott, Science 277 (1997) 1799.

[6] D. Tsiulyanu, S. Marian, H.D. Liess, Sensors and Actuators B: Chemical 85 (2002) 232.

[7] V. Georgieva, T. Yordanov, V. Pamukchieva, D. Arsova, V. Gadjanova, L. Vergov, AIP Conference Proceedings 1203 (2010) 1079.

[8] F. Ménil, V. Coillard, C. Lucat, Sensors and Actuators B: Chemical 67 (2000) 1.

[9] M. Penza, C. Martucci, G. Cassano, Sensors and Actuators B: Chemical 50 (1998) 52.

[10] G. Sberveglieri, L. Depero, S. Groppelli, P. Nelli, Sensors and Actuators B: Chemical 26 (1995) 89.

[11] M. Ferroni, V. Guidi, G. Martinelli, M. Sacerdoti, P. Nelli, G. Sberveglieri, Sensors and Actuators B: Chemical 48 (1998) 285.

[12] A.K. Sharma, R. Jha, J. Appl. Phys. 106 (2009) 103101.

[13] V. Georgieva, Z. Raicheva, A. Grechnikov, V. Gadjanova, M. Atanassov, J. Lazarov, E. Manolov, Journal of Physics: Conference Series 253 (2010) 012046.

[14] A.K. Prasad, P.I. Gouma, J. Mater. Sci. 38 (2003) 4347.

[15] V. Georgieva, M. Mitkova, P. Chen, D. Tenne, K. Wolf, V. Gadjanova, Mater. Chem. Phys. (submitted).

[16] R. Shannon, Acta Crystallographica Section A 32 (1976) 751.

[17] J.E. Huheey, E.A. Keiter, R.L. Keiter, Inorganic chemistry: principles of structure and reactivity, Pearson Education, 2000.

[18] Z.U. Borisova, Glassy semiconductors, Plenum Press, New York, 1981.

[19] A. Feltz, Amorphous inorganic materials and glasses, VCH, 1993.

[20] P. Boolchand, P. Chen, U. Vempati, J. Non-Cryst. Sol. 355 (2009) 1773.

[21] V. Boev, M. Mitkova, E. Lefterova, T. Wagner, S. Kasap, M. Vlcek, J. Non-Cryst. Sol. 266-269 (2000) 867.

[22] K.J. Rao, Structural chemistry of glasses, Elsevier, 2002.

[23] M.F. Thorpe, personal communication.

[24] M.D. Ingram, M.A. Mackenzie, W. Müller, M. Torge, Solid State Ionics 28-30 (1988) 677.

[25] J.C. Phillips, J. Non-Cryst. Sol. 34 (1979) 153.

[26] X.W. Feng, W.J. Bresser, P. Boolchand, Phys. Rev. Lett. 78 (1997) 4422.

[27] P. Boolchand, M. Micoulaut, P. Chen, in: S. Raoux, M. Wuttig (Eds.), Phase Change Materials: Science and Applications, Springer, Heidelberg, 2008, p. 37.

[28] S.R. Elliott, J. Non-Cryst. Sol. 182 (1995) 40.

[29] J. Blétry, Phil. Mag. B 62 (1990) 469.

[30] J. Haubrich, R.G. Quiller, L. Benz, Z. Liu, C.M. Friend, Langmuir 26 (2010) 2445.

[31] G. Sauerbrey, Z. Phys. A: Hadrons Nucl. 155 (1959) 206.

[32] M. Kastner, H. Fritzsche, Phil. Mag. B 37 (1978) 199.

[33] http://en.wikipedia.org/wiki/Nitrogen_dioxide.

[34] M. Cobb, D.A. Drabold, R.L. Cappelletti, Phys. Rev. B 54 (1996) 12162.

[35] M. Cobb, D.A. Drabold, Phys. Rev. B 56 (1997) 3054.

**Figure captions**

**Figure 1.** Packing fraction as a function of the composition for binary chalcogenide glasses and some Ag-containing chalco-halide chalcogenide glasses.

**Figure 2.** The gas absorption experimental setup.

**Figure 3.** Deconvoluted Raman spectra of 200nm-thick deposited $Ge_{20}Se_{80}$ glass films before and after gas absorption.

**Figure 4.** XRD data for the $Ge_{20}Se_{80}$ films, the XRD intensity I(Q) was plotted against scattering vector Q (= $4\pi\sin\theta/\lambda$).

**Figure 5.** Ge 3d core level XPS spectra for $Ge_{20}Se_{80}$ thin film (200nm thickness, resonator 5-38). a) as-deposited; b) after $NO_2$ gas absorption.

**Figure 6.** N 1s XPS spectra for $Ge_{20}Se_{80}$ thin film (200nm thickness, resonator 5-38). a) as-deposited; b) after $NO_2$ gas absorption.

**Figure 7.** AFM data for $Ge_{20}Se_{80}$ thin film (200nm thickness, sample# 5-38 / 040411). a) as-deposited; b) after $NO_2$ gas absorption.

**Figure 8.** Frequency – time characteristic (FTC) of $Ge_{20}Se_{80}$-QCM at 500 ppm NO2 concentration. Thickness of $Ge_{20}Se_{80}$ – 199nm

**Figure 9.** Frequency – time characteristic (FTC) of $Ge_{20}Se_{80}$-QCM system at different $NO_2$ concentrations. Thickness of $Ge_{20}Se_{80}$– 199nm

**Figure 10.** Dependence of the $Ge_{20}Se_{80}$-QCM frequency shift and sorbed mass towards $NO_2$ concentrations at different thickness of the sensitive layer. (SAMPLES: 5-34, 5-35, 5-38)

**Figure 11**. Structure of the $NO_2$ molecule [33].

**Figure 12.** Model of the structure of the $Ge_{20}Se_{80}$ glass with the $NO_2$ molecule diffusing in it.

**Table captions**

**Table 1.** EDS composition for Ge-Se thin films

**Table 1.**

| Sample name | Source Composition | Thickness | EDS Ge% | EDS Se% | Standard deviation |
|---|---|---|---|---|---|
| Resonators 5-31 & 5-32 | $Ge_{20}Se_{80}$ | 50.9 nm | 20.50 | 79.50 | 0.56 |
| | $Ge_{20}Se_{80}$ | 55.1 nm | 19.23 | 80.77 | 0.96 |
| Resonators 5-33 & 5-34 | $Ge_{20}Se_{80}$ | 101 nm | 19.65 | 80.36 | 1.27 |
| | $Ge_{20}Se_{80}$ | 101 nm | 20.70 | 79.30 | 0.03 |
| Resonators 5-35 & 5-36 | $Ge_{20}Se_{80}$ | 150 nm | 21.52 | 78.49 | 0.40 |
| | $Ge_{20}Se_{80}$ | 151 nm | 20.93 | 79.07 | 0.00 |
| Resonators 5-37 & 5-38 | $Ge_{20}Se_{80}$ | 198.7 nm | 20.71 | 79.29 | 0.66 |
| | $Ge_{20}Se_{80}$ | 199.3 nm | 20.83 | 79.17 | 0.79 |

**Figure 1.**

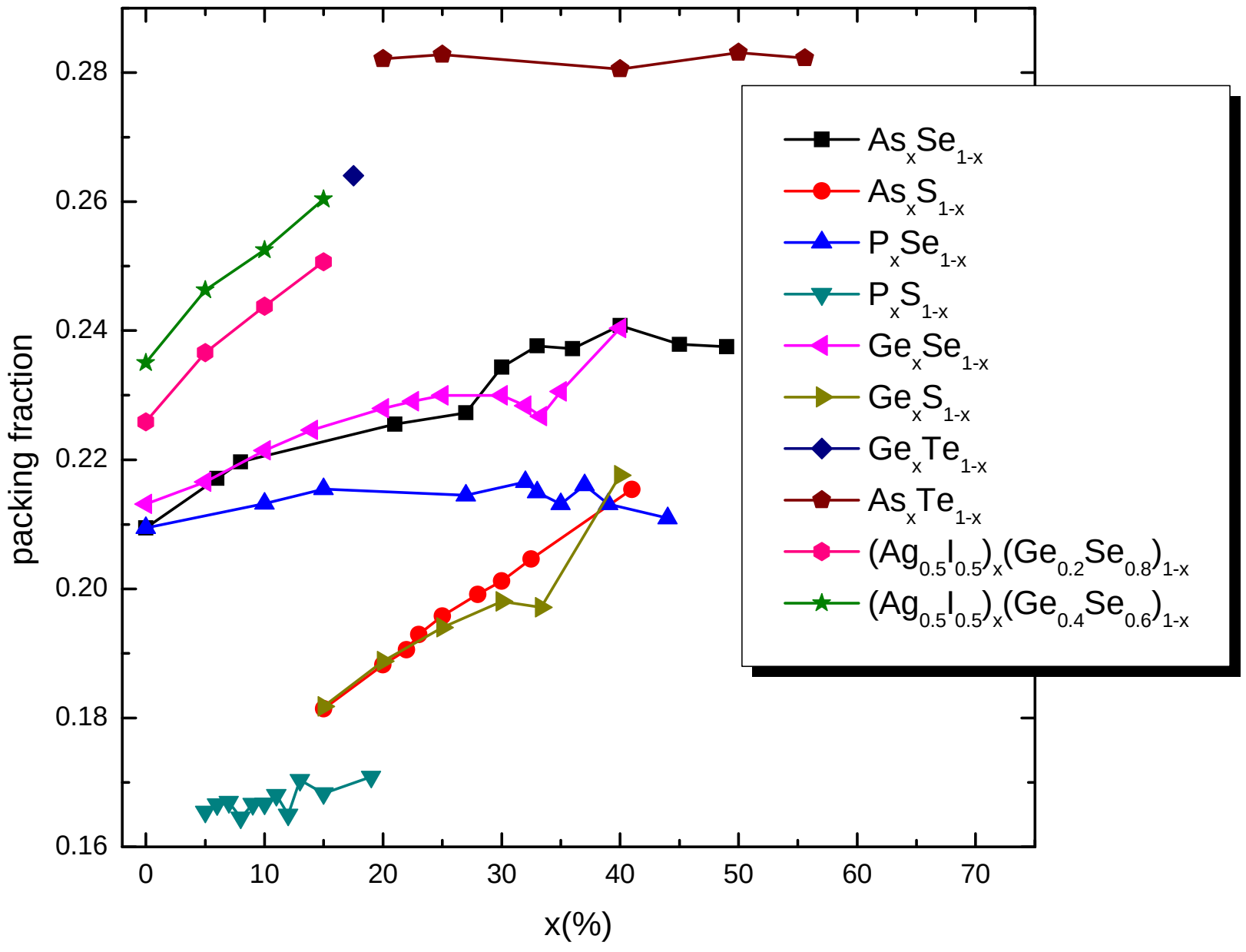

**Figure 2.**

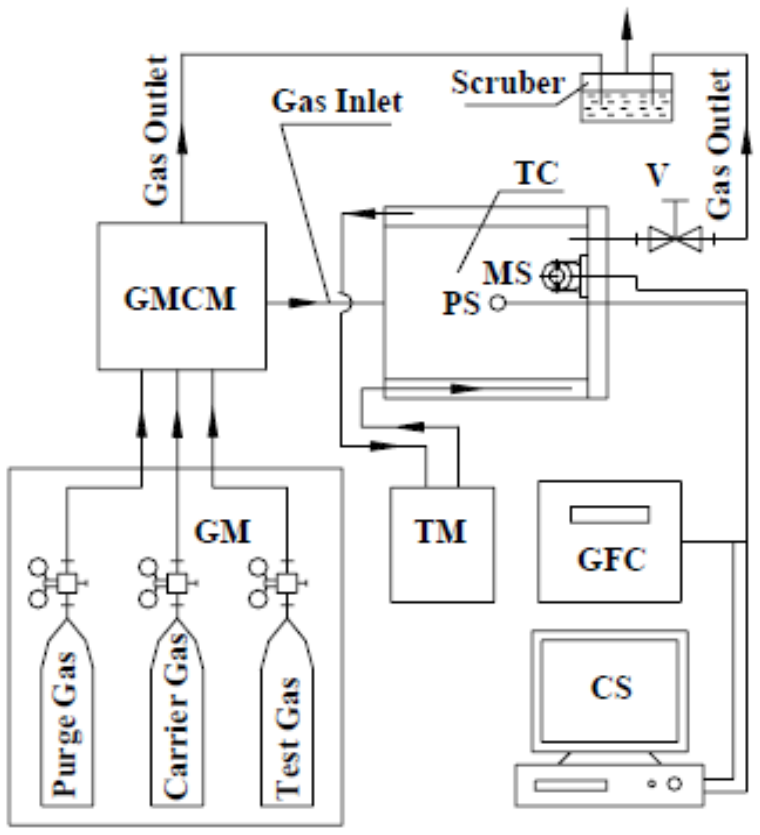

**Figure 3.**

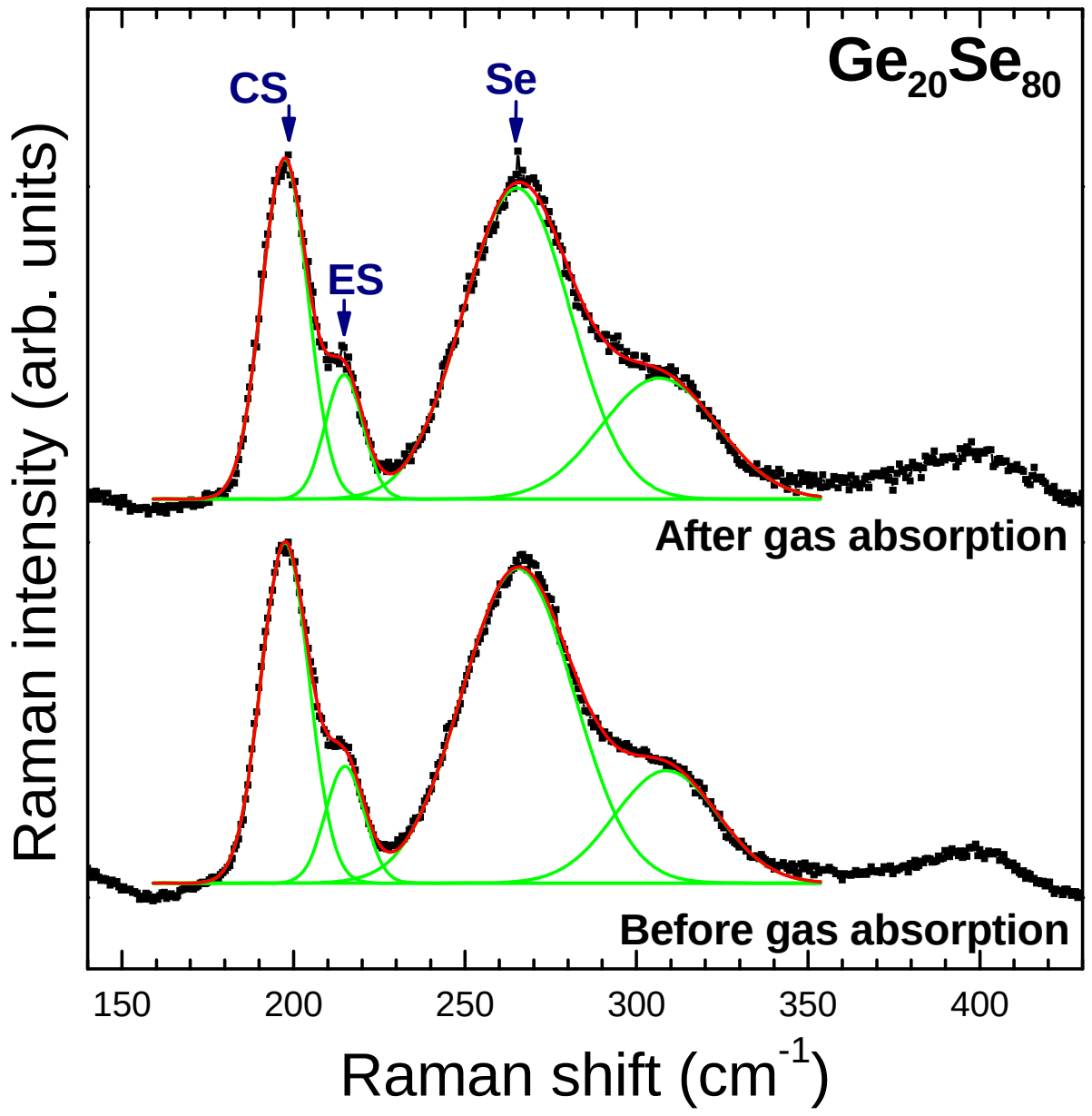

**Figure 4.**

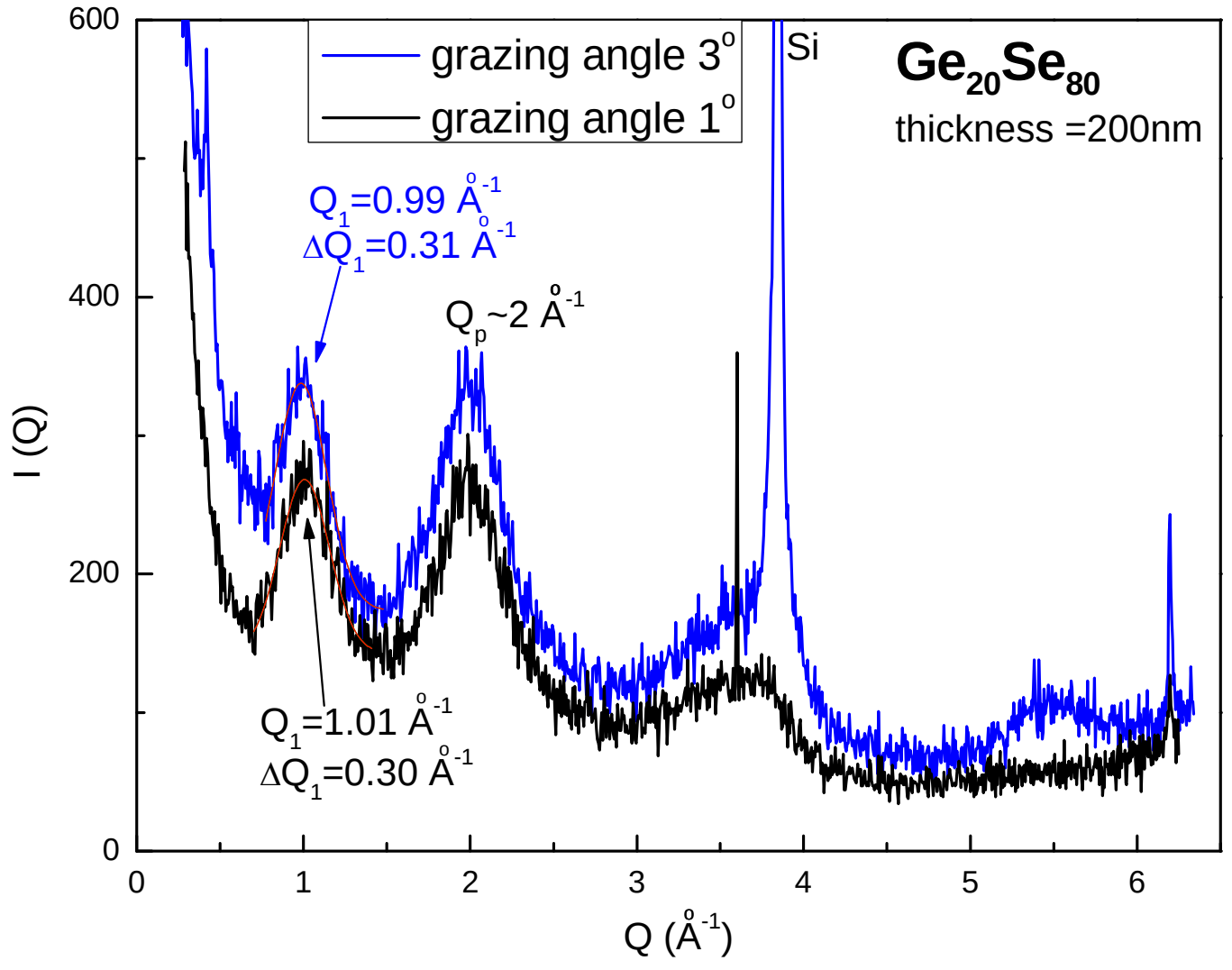

**Figure 5.**

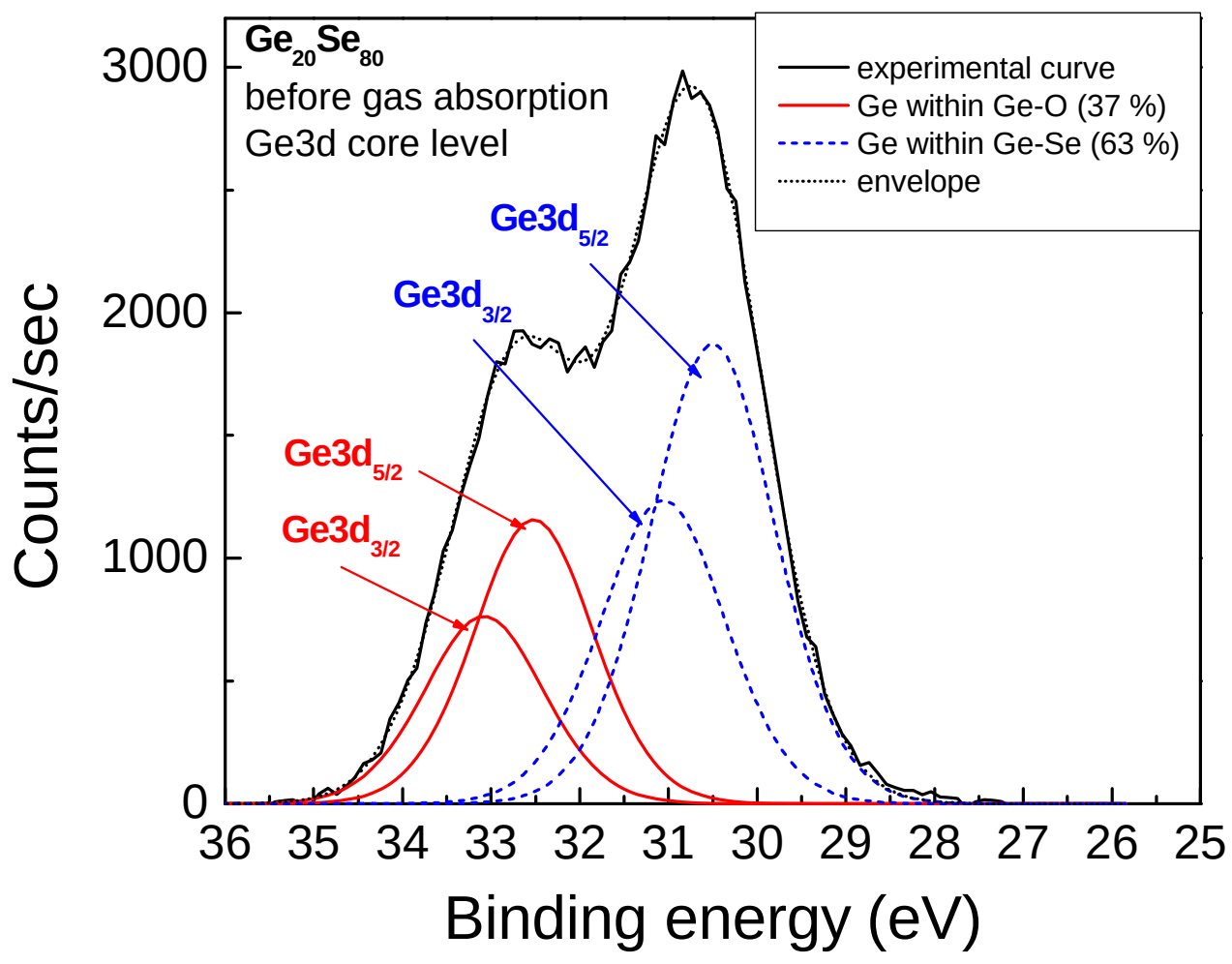
Ge20Se80
before gas absorption
Ge3d core level
experimental curve
Ge within Ge-O (37 %)
Ge within Ge-Se (63 %)
envelope
Ge3d5/2
Ge3d3/2
Ge3d5/2
Ge3d3/2
Counts/sec
0
1000
2000
3000
36 35 34 33 32 31 30 29 28 27 26 25
Binding energy (eV)

(a)

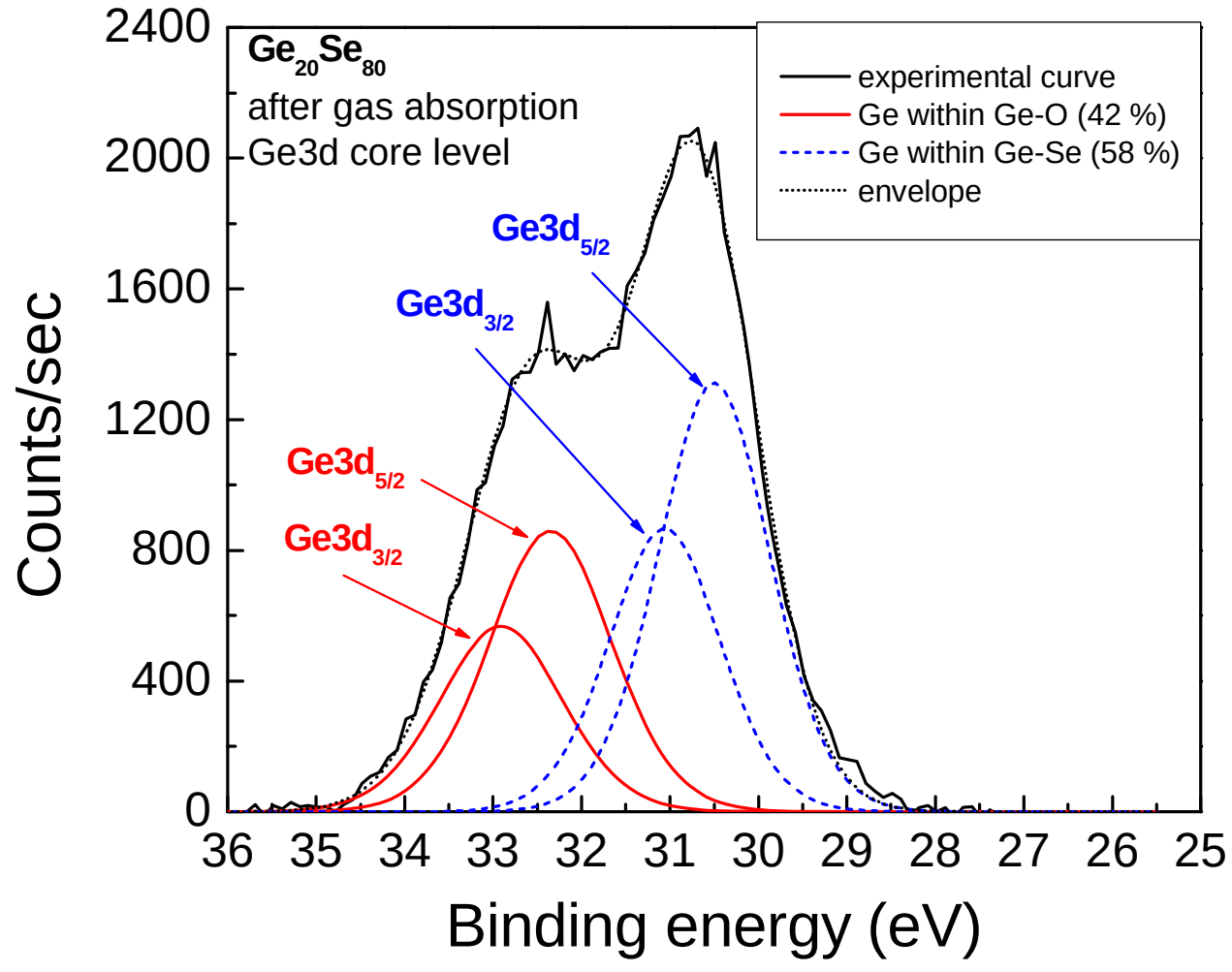
Ge20Se80
after gas absorption
Ge3d core level
experimental curve
Ge within Ge-O (42 %)
Ge within Ge-Se (58 %)
envelope
Ge3d5/2
Ge3d3/2
Ge3d5/2
Ge3d3/2
Counts/sec
0
400
800
1200
1600
2000
2400
36 35 34 33 32 31 30 29 28 27 26 25
Binding energy (eV)

(b)

**Figure 6.**

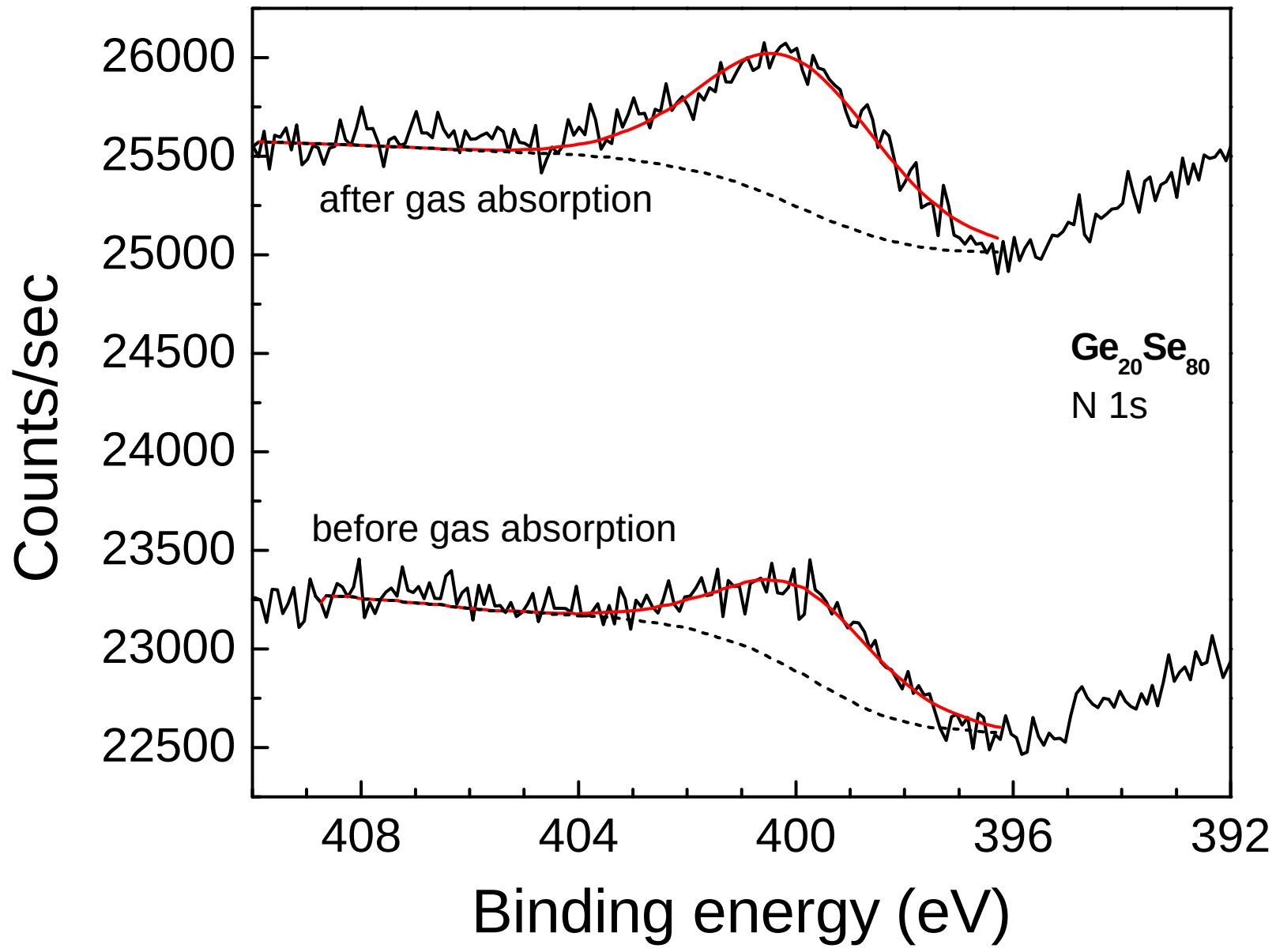

**Figure 7.**

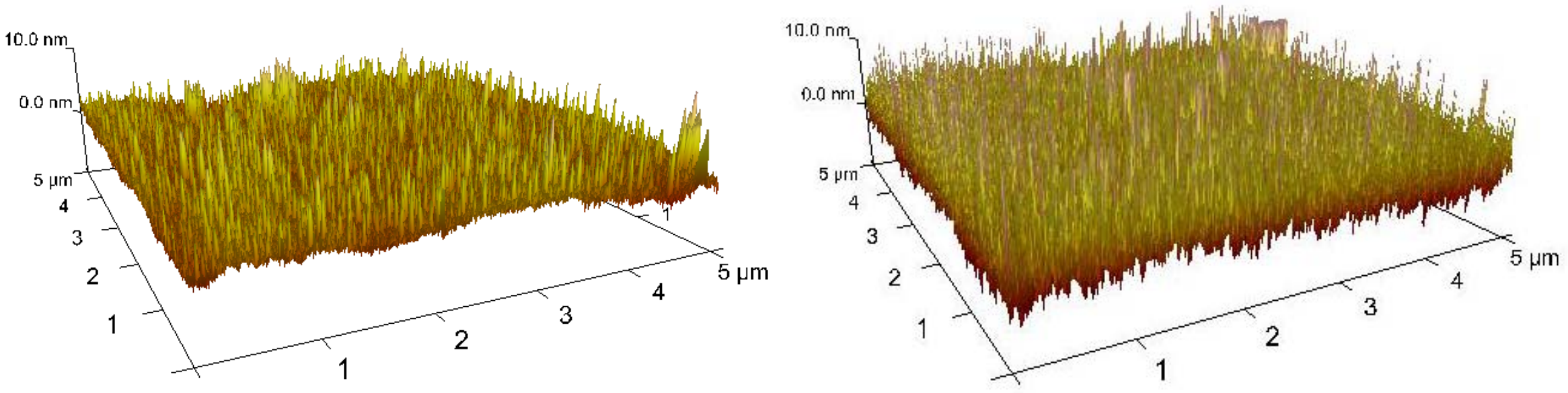

(a) (b)

**Figure 8.**

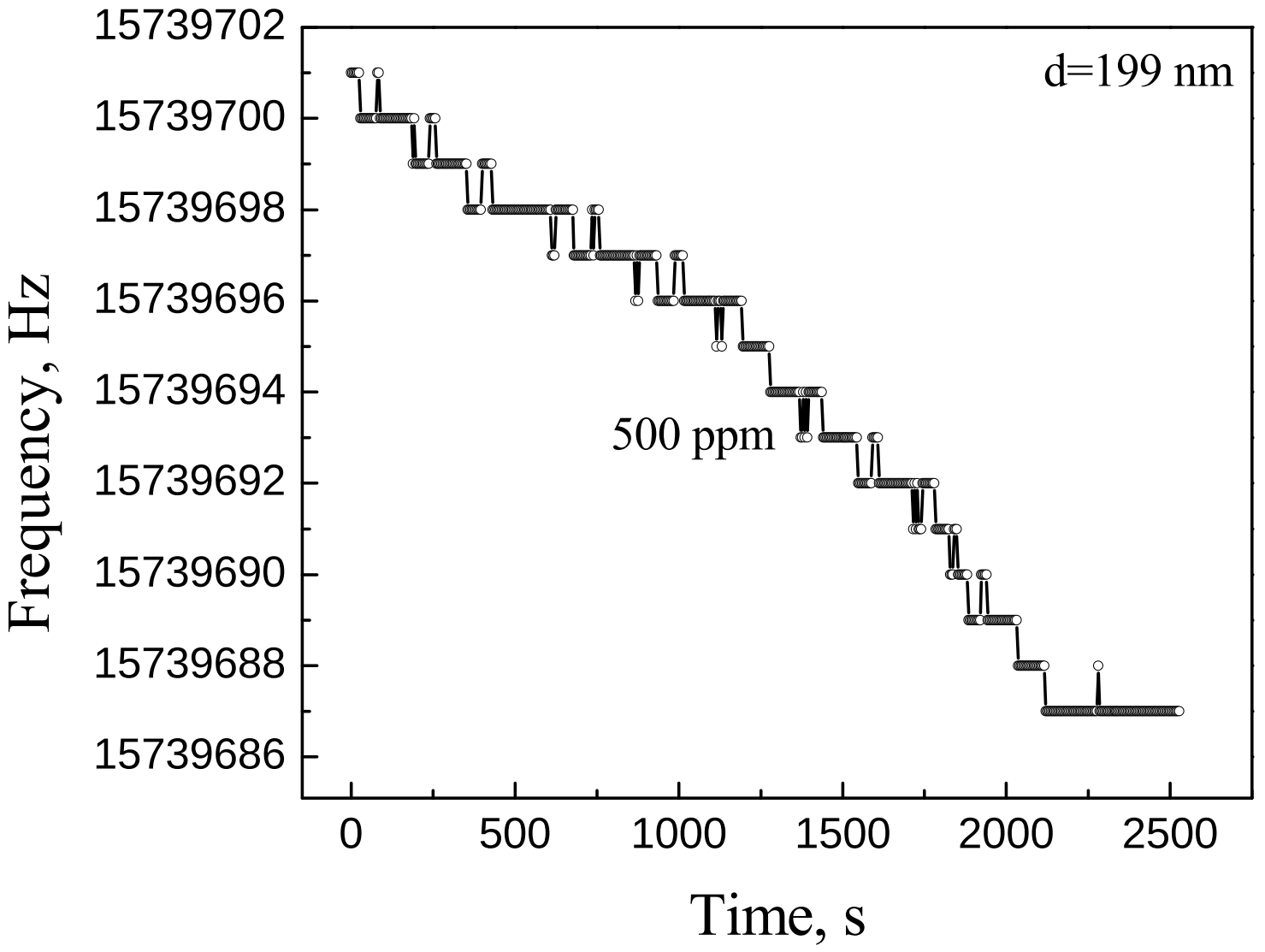

**Figure 9.**

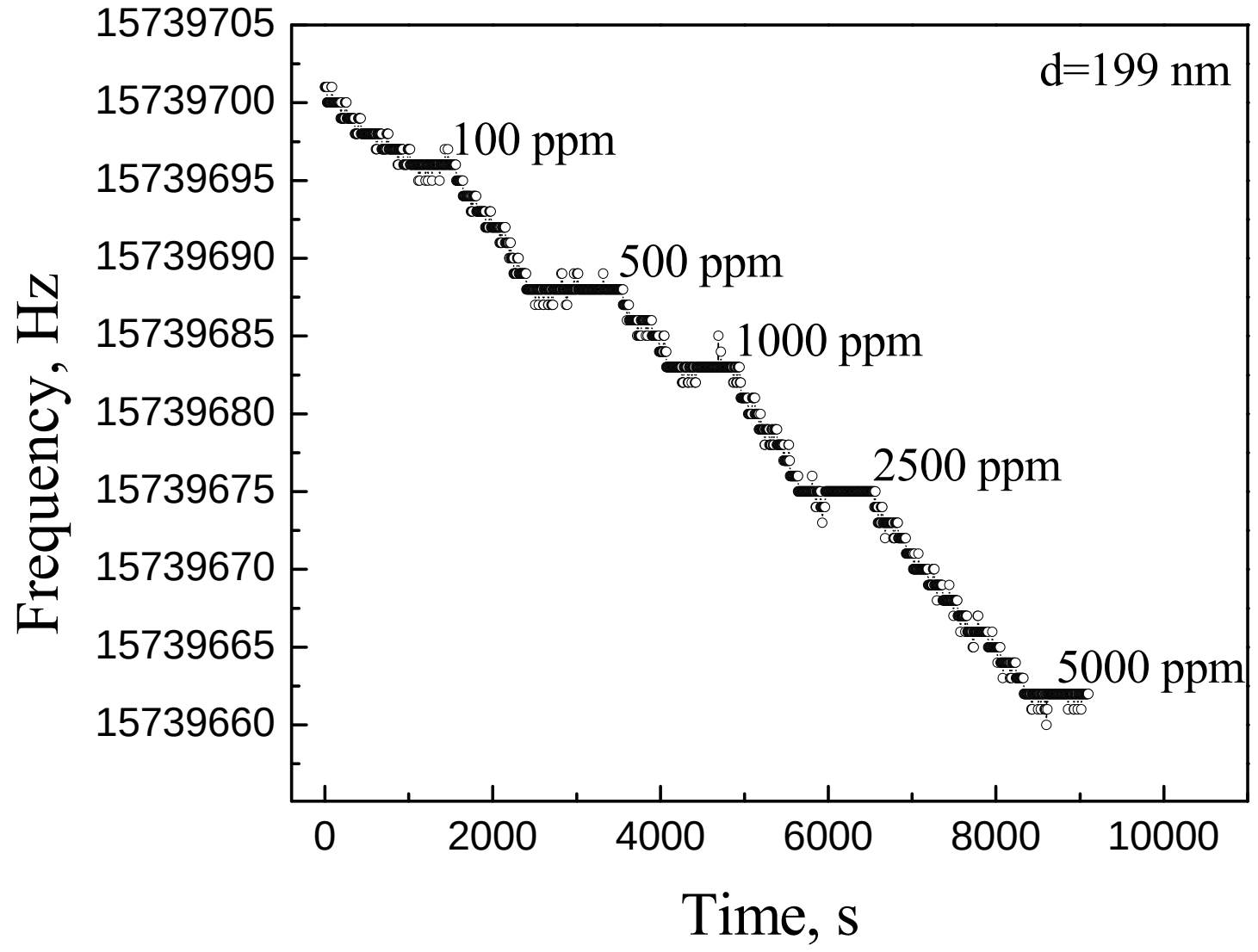

**Figure 10.**

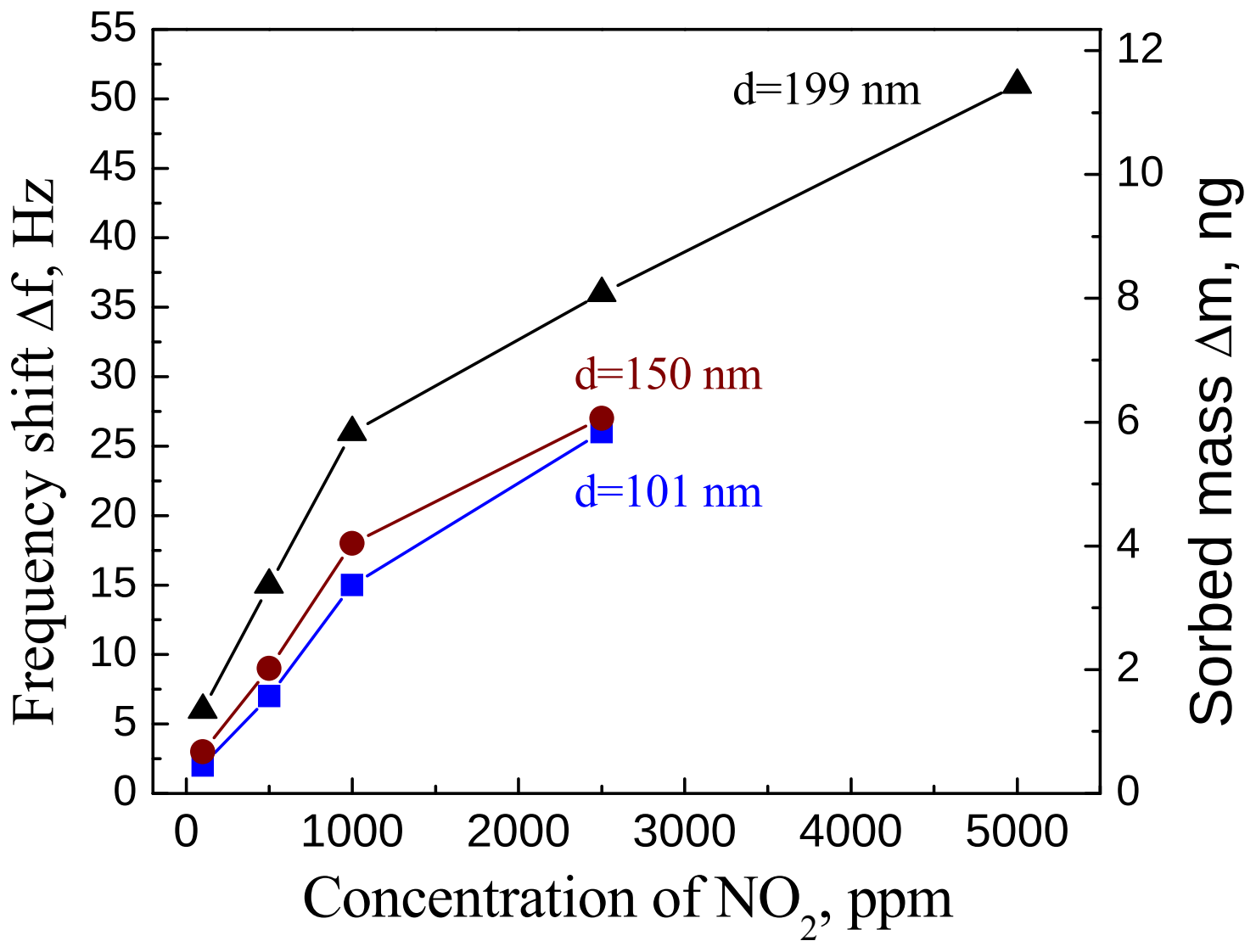

**Figure 11.**

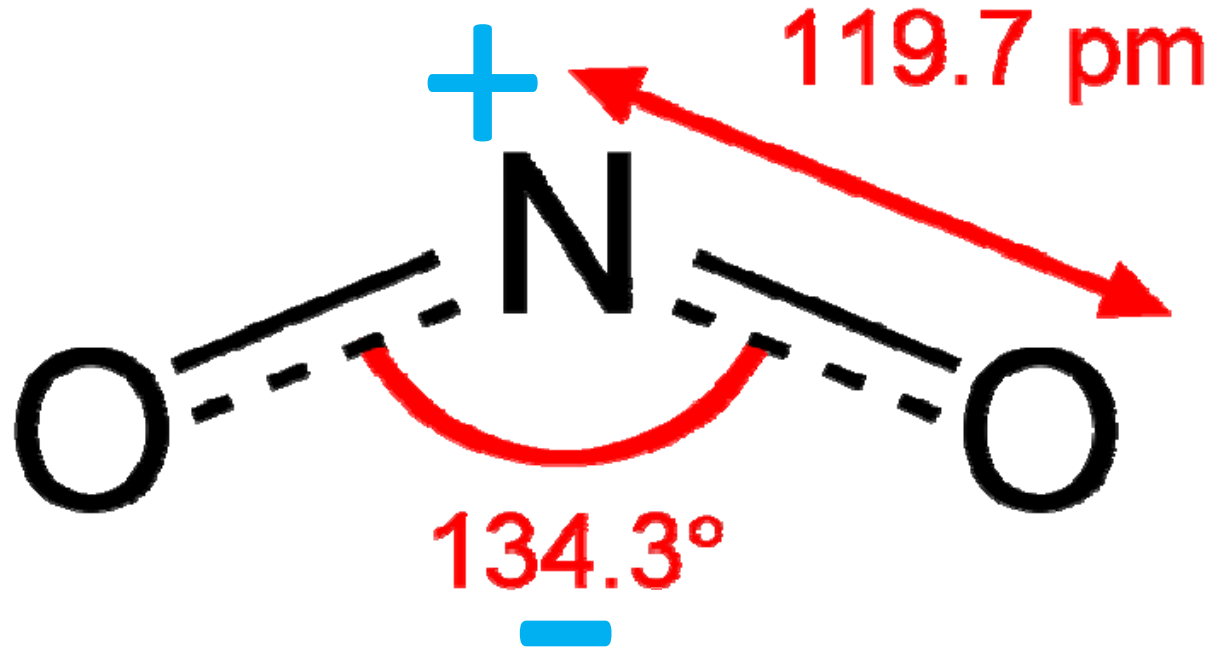

**Figure 12.**

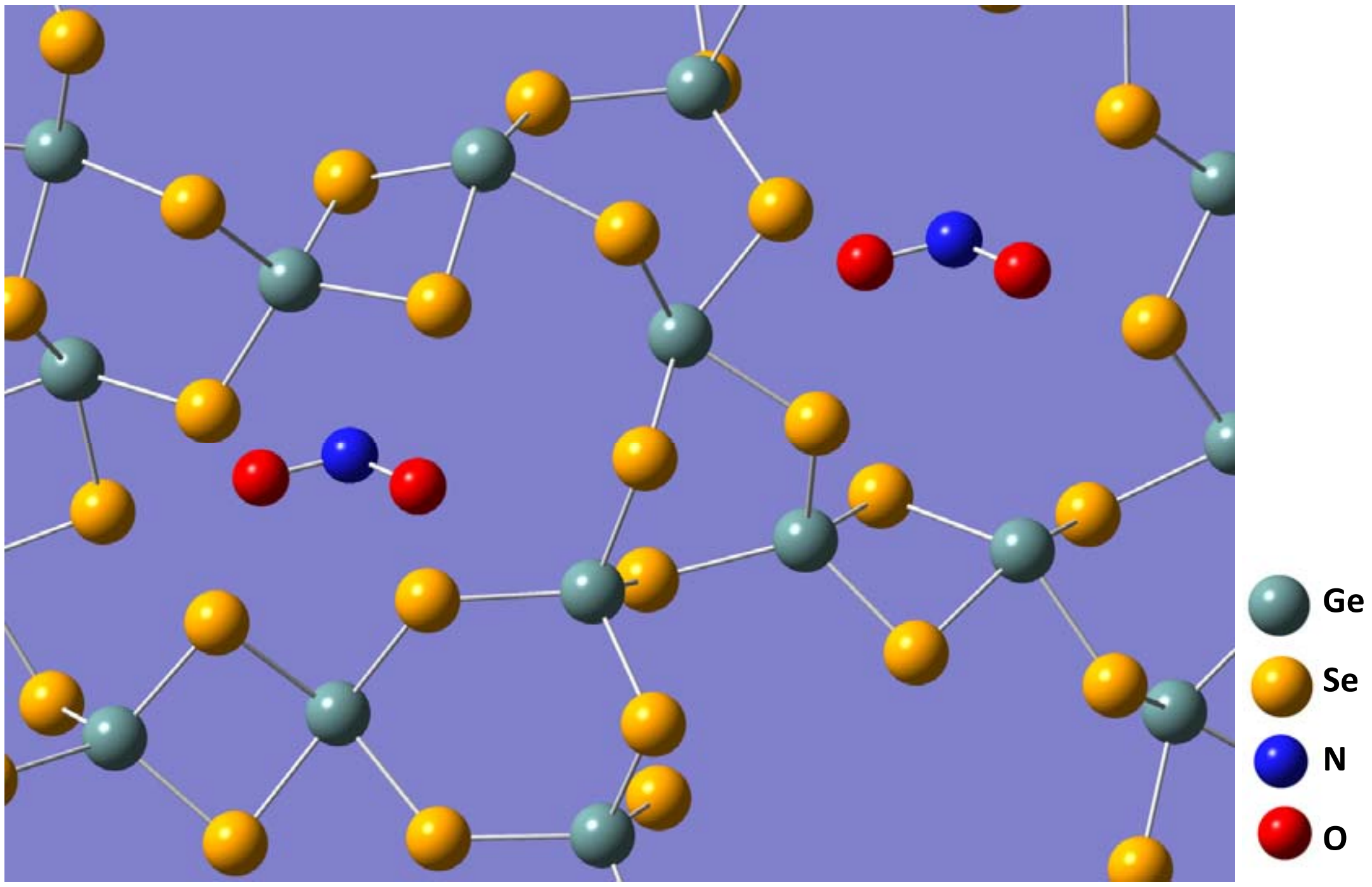